\documentclass[12pt]{iopart}

\usepackage{iopams}  
\usepackage{graphics}
\begin{document}

\title{B\"acklund transforms of the extreme Kerr
near-horizon geometry}

\author{Reinhard Meinel and Andreas Kleinw\"achter}

\address{Theoretisch-Physikalisches Institut,
University of Jena,\\
 Max-Wien-Platz 1, 07743 Jena, Germany}

\ead{meinel@tpi.uni-jena.de, kleinwaechter@tpi.uni-jena.de}

\begin{abstract}
We apply the method of B\"acklund transformations to generate a new hierarchy of exact solutions to 
the vacuum Einstein equations starting from the extreme Kerr near-horizon geometry. Solutions with extreme 
Kerr near-horizon asymptotics containing an arbitrary number of free parameters are included.
\end{abstract}

\vspace{5pc}

{\it Dedicated to Gernot Neugebauer on the occasion of his 80th birthday}

\vspace{8pc}

\section{Introduction}
The general relativistic solution describing a uniformly rotating disk of dust \cite{nm95} admits an interesting 
parameter limit leading to the extreme Kerr metric outside the horizon (``exterior perspective''). 
From the ``interior perspective'', the same parameter limit can be performed after an appropriate parameter dependent coordinate 
transformation and then leads to a spacetime that still describes a rotating disk of dust embedded in a non-asymptotically flat 
vacuum exterior that approaches the extreme Kerr near-horizon geometry at spatial infinity \cite{meinel97}, see also \cite{rfe}. 
The same phenomenon was observed numerically for ``relativistic Dyson rings'' \cite{akm03}. Moreover, it was shown that a “black hole limit” of 
rotating fluid bodies in equilibrium always leads to the extreme Kerr solution \cite{meinel06}. 

This motivates the systematic investigation of solutions
which have the extreme Kerr near-horizon geometry, also called ``extreme Kerr throat geometry'' \cite{bh99}, as their asymptotics. 
As a first step, we derive here a new class of solutions by means of B\"acklund transformations\footnote{The method of B\"acklund 
transformations is closely related to other soliton-theoretic techniques for solving certain nonlinear partial differential equations and can be applied to the vacuum Einstein equations in the presence of 
two commuting Killing vectors, see \cite{nmpz, bv01, rs02}.} \cite{harrison78, neugebauer79} using  
the extreme Kerr near-horizon geometry as the starting point (``seed solution'') and show how the 
B\"acklund parameters have to be constrained in order to conserve the extreme Kerr near-horizon asymptotics.  
Of course, the solutions constructed in this way are pure vacuum solutions. Instead of material sources they possess some singularities; 
note that this guarantees consistency with results published in \cite{ahmr09, drs09}. However, by removing a portion of the solution that 
contains the singularities and replacing it with an appropriate non-vacuum solution, a meaningful, globally regular spacetime with 
extreme Kerr near-horizon asymptotics could be obtained. 

The derived solutions might also be of interest within the context of the Kerr/CFT correspondence, see \cite{compere17}.
%%%%%%%%%%%%%%%%%%%%%%%%%%%%%%%%%%%%%%%%%%%%%%%%%%%%%%%%%%%%%%%%%%%%%%%
\section{Ernst potential of the extreme Kerr near-horizon geometry}
It is well known that the stationary and axially symmetric vacuum Einstein equations are equivalent to the Ernst equation \cite{ernst68, kn68}
\begin{equation}\label{ernst}
(\Re\, {\mathcal E})\nabla^2 {\mathcal E}=(\nabla {\mathcal E})^2,
\end{equation}
where the operator $\nabla$ has the same meaning as in Euclidean 3-space in
which $r$, $\theta$ and $\phi$ are spherical coordinates. The complex Ernst potential $\mathcal E$ 
depends on $r$ and $\theta$ only. The spacetime line element reads
\begin{equation}\label{line}
{\rm d}s^2=f^{-1}[\,h({\rm d}r^2+r^2{\rm d}\theta^2)+r^2\sin^2\theta\,{\rm d}\phi^2]
-f({\rm d}t+a\,{\rm d}\phi)^2
\end{equation}
with $f=\Re\, {\mathcal E}$; the other metric functions $h$ and $a$ can also be obtained from $\mathcal E$.
The Ernst potential of the extreme Kerr near-horizon geometry is given by \cite{meinel97}
\begin{equation}
 \mathcal E_{\rm NHG} = -\Omega^2r^2H(\theta), \quad H(\theta)=
 \frac{2(1+{\rm i}\cos\theta)^2}{1-{\rm i}\cos\theta}+\sin^2\theta.
 \label{EP}
\end{equation}
The real constant $\Omega$ is the angular velocity of the horizon.  Note that (\ref{EP})  belongs to a family of 
solutions discovered by Ernst \cite{ernst77}.
%%%%%%%%%%%%%%%%%%%%%%%%%%%%%%%%%%%%%%%%%%%%%%%%%%%%%%%%%%%%%%%%%%%%%%%
\section{B\"acklund transformation}
Neugebauer's general B\"acklund formula \cite{neugebauer80a} has been used to investigate huge classes of asymptotically 
flat solutions to the stationary and axially symmetric vacuum Einstein equations. However, it can be applied to 
non-asymptotically flat solutions as well. In general, for a given seed solution $\mathcal E_0$, the ($2n$-fold) 
B\"acklund transform $\mathcal E$ reads \cite{neugebauer80a, neugebauer80b}
\medskip
\begin{equation}
 \mathcal E =  \mathcal E_0\; \frac{\det\left(\begin{array}{ccccccc}
1 & 1 & 1 & \cdot & \cdot & \cdot & 1\\
\alpha_0 & \alpha_1\lambda_1 & \alpha_2\lambda_2 & \cdot & \cdot & \cdot & \alpha_{2n}\lambda_{2n} \\
1 & (\lambda_1)^2 & (\lambda_2)^2 & \cdot & \cdot & \cdot & (\lambda_{2n})^2 \\
\alpha_0 & \alpha_1(\lambda_1)^3 & \alpha_2(\lambda_2)^3 &  \cdot & \cdot & \cdot & \alpha_{2n}(\lambda_{2n})^3 \\
1 & (\lambda_1)^4 & (\lambda_2)^4 & \cdot & \cdot & \cdot & (\lambda_{2n})^4 \\
\alpha_0 & \alpha_1(\lambda_1)^5 & \alpha_2(\lambda_2)^5 & \cdot & \cdot & \cdot & \alpha_{2n}(\lambda_{2n})^5 \\
\cdot & \cdot & \cdot &  \cdot & \cdot& \cdot & \cdot \\
\cdot & \cdot & \cdot & \cdot & \cdot & \cdot & \cdot \\
1 &  (\lambda_1)^{2n} & (\lambda_2)^{2n} & \cdot & \cdot & \cdot & (\lambda_{2n})^{2n}
\end{array}\right)}{\det\left(\begin{array}{ccccccc}
1 & 1 & 1 & \cdot & \cdot & \cdot & 1\\
1 & \alpha_1\lambda_1 & \alpha_2\lambda_2 & \cdot & \cdot & \cdot & \alpha_{2n}\lambda_{2n} \\
1 & (\lambda_1)^2 & (\lambda_2)^2 & \cdot & \cdot & \cdot & (\lambda_{2n})^2 \\
1 & \alpha_1(\lambda_1)^3 & \alpha_2(\lambda_2)^3 &  \cdot & \cdot & \cdot & \alpha_{2n}(\lambda_{2n})^3 \\
1 & (\lambda_1)^4 & (\lambda_2)^4 & \cdot & \cdot & \cdot & (\lambda_{2n})^4 \\
1 & \alpha_1(\lambda_1)^5 & \alpha_2(\lambda_2)^5 & \cdot & \cdot & \cdot & \alpha_{2n}(\lambda_{2n})^5 \\
\cdot & \cdot & \cdot &  \cdot & \cdot& \cdot & \cdot \\
\cdot & \cdot & \cdot & \cdot & \cdot & \cdot & \cdot \\
1 &  (\lambda_1)^{2n} & (\lambda_2)^{2n} & \cdot & \cdot & \cdot & (\lambda_{2n})^{2n}
\end{array}\right)},
\label{det}
\end{equation}
where the entries in the two $(2n+1)\times(2n+1)$ matrices are given by
\begin{equation}
\alpha_0=-\frac{\mathcal E_0^*}{\mathcal E_0}, \quad \lambda_i=
\sqrt{\frac{K_i-r{\rm e}^{{\rm i}\theta}}{K_i-r{\rm e}^{-{\rm i}\theta}}} \quad (\lambda_i\to{\rm e}^{{\rm i}\theta}\; \mbox{as} \; 
r\to\infty)
\label{lambda}
\end{equation}
and solutions $\alpha_i$ to the total Riccati equations
\begin{equation}
 \begin{array}{ccl}
  (\mathcal E_0 +
  \mathcal E^*_0)\,\mathrm d \alpha_i & = & \left[\frac{\partial\mathcal E_0^*}{\partial z}(\alpha_i-\lambda_i) + 
  \frac{\partial\mathcal E_0}{\partial z}\alpha_i(\alpha_i\lambda_i-1)\right] \mathrm d z \\
  & & + \; \left[\frac{\partial \mathcal E_0^*}{\partial z^*}(\alpha_i-\lambda_i^{-1}) + 
  \frac{\partial \mathcal E_0}{\partial z^*}\alpha_i(\alpha_i\lambda_i^{-1}-1)\right] \mathrm d z^* 
 \end{array}
 \label{Ricc}
\end{equation}
with the complex coordinates
\begin{equation}
 z={\mathrm i}r{\rm e}^{-\mathrm i \theta}, \quad z^*=-{\mathrm i}r{\rm e}^{\mathrm i \theta}.
\end{equation}
Note that the integrability condition of (\ref{Ricc}) is equivalent to the Ernst equation for $\mathcal E_0$. 
Via a quotient ansatz, the Riccati equation can 
be reformulated as a system of linear differential equations \cite{neugebauer80b}. 
The asymptotic fixing of $\lambda_i$ as given in (\ref{lambda}) can be chosen without loss of 
generality since the $\lambda_i$'s enter (\ref{det}) only with even powers and as products $\alpha_i\lambda_i$. 
To any solution $\alpha_i$ of the Riccati equation (\ref{Ricc}) with $\lambda_i$ one has a corresponding 
solution $-\alpha_i$ with $-\lambda_i$.

The (finite) constants $K_i$ must either be real ($K_i=K_i^*$), with the consequence $\lambda_i=1/\lambda_i^*$, 
or complex conjugate pairs ($K_j=K_i^*$), to ensure 
$\lambda_j=1/\lambda_i^*$. The integration constants of the Riccati equations (\ref{Ricc})
have to be chosen such that $\alpha_i=1/\alpha_i^*$ or $\alpha_j=1/\alpha_i^*$, respectively. 

For the particular seed solution (\ref{EP}) we obtain
\begin{equation}
 \alpha_0=-\frac{H^*}{H}
 \label{al0}
\end{equation}
and
\begin{equation}
 \alpha_i=
 -\frac{\psi(\lambda_i,\theta)-c_i\psi(-\lambda_i,\theta)}{\chi(\lambda_i,\theta)+c_i\chi(-\lambda_i,\theta)}
 \qquad (i=1,2,\dots,2n)
 \label{ali}
\end{equation}
with
\begin{equation}
\psi(\lambda,\theta)=[\chi(1/\lambda^*,\theta)]^*=
\frac{A(1+\lambda^2)+B(1-\lambda^2)+C\lambda}{{\rm e}^{-{\rm i}\theta}(\lambda-{\rm e}^{{\rm i}\theta})^2},
\end{equation}
\begin{equation}
 A=\frac{\cos\theta+{\rm i}}{1+{\rm i}}, \quad B=\frac{(1-{\rm i})\sin\theta}{\cos\theta-{\rm i}}
\end{equation}
and 
\begin{equation}
  C=-(1+{\rm i})(\cos\theta-{\rm i}).  
\label{C}  
\end{equation}
The constants $c_i$, which can also be chosen infinite [meaning $\alpha_i=\psi(-\lambda_i,\theta)/\chi(-\lambda_i,\theta)$], have to satisfy 
\begin{equation}
 c_i=-c_i^* \quad \mbox{(for real $K_i$)} \quad \mbox{or}
 \quad c_j=-c_i^*\quad \mbox{(for pairs $K_j=K_i^*$)}.
 \label{condci}
\end{equation}

\bigskip

\noindent We mention that $\alpha_0$ solves the Riccati equation (\ref{Ricc}) with $\lambda_i$ replaced by $1$ and can be expressed here as
\begin{equation}
 \alpha_0=-\lim\limits_{\lambda\to1}\frac{\psi(\lambda,\theta)+{\rm i}\psi(-\lambda,\theta)}{\chi(\lambda,\theta)-
 {\rm i}\chi(-\lambda,\theta)}
\end{equation}
(note that 
$\psi(1,\theta)=-\chi(1,\theta)=-{\rm i}$, 
$\psi(-1,\theta)=\chi(-1,\theta)=1$). 
%%%%%%%%%%%%%%%%%%%%%%%%%%%%%%%%%%%%%%%%%%%%%%%%%%%%%%%%%%%%%%%%%%%%%%%
\section{Solutions with extreme Kerr near-horizon asymptotics}
For the discussion of the asymptotic behaviour as $r\to\infty$ 
a reformulation of (\ref{det}) in terms of $n \times n$ determinants following \cite{yamazaki83} is useful, see also \cite{mns91}. With
\begin{equation}
 r_i \equiv -\lambda_i(K_i-r{\rm e}^{-{\rm i}\theta})
 =r\sqrt{\left(1-\frac{K_i{\rm e}^{{\rm i}\theta}}{r}\right)
 \left(1-\frac{K_i{\rm e}^{-{\rm i}\theta}}{r}\right)}
 \label{ri}
\end{equation}
one obtains 
\begin{equation}
 \mathcal E =  \mathcal E_0\; \frac{\det\left(\frac{\alpha_p r_p-\alpha_q r_q}{K_p-K_q}+\alpha_0\right)}
 {\det\left(\frac{\alpha_p r_p-\alpha_q r_q}{K_p-K_q}+1\right)}
 \label{detn}
\end{equation}
with
\begin{equation}
 p=1,3,5,\dots, 2n-1; \quad q=2,4,6,\dots, 2n.
\end{equation}
(This means: first row $p=1$, second row $p=3$, $\dots$, $n$-th row $p=2n-1$ and 
first column $q=2$, second column $q=4$, $\dots$, $n$-th column $q=2n$.) For $n=1$, Eq.~(\ref{detn}) reduces to
\begin{equation}
 \mathcal E =  \mathcal E_0\; \frac{\alpha_1 r_1-\alpha_2 r_2+\alpha_0(K_1-K_2)}
 {\alpha_1 r_1-\alpha_2 r_2+K_1-K_2}\,.
\end{equation}

By means of an expansion in powers of $r^{-1}$ it can easily be verified from (\ref{lambda}) and (\ref{ali})--(\ref{C}) that 
\begin{equation}
 \alpha_i=-\frac{\psi(\lambda_i,\theta)}{\chi(\lambda_i,\theta)} + \mathcal O (r^{-2})=F(\theta) + \mathcal O (r^{-1}) \quad \mbox{for $c_i\neq\infty$}
\end{equation}
and 
\begin{equation}
 \alpha_i=\frac{\psi(-\lambda_i,\theta)}{\chi(-\lambda_i,\theta)}=G(\theta) + \mathcal O (r^{-1}) \quad \mbox{for $c_i=\infty$}
\end{equation}
with 
\begin{equation}
 F(\theta)=\frac{{\rm i}(\cos\theta+{\rm i})^2}{(\cos\theta-{\rm i})^2}
\end{equation}
and 
\begin{equation}
 G(\theta)=\frac{{\rm i}(\cos\theta+{\rm i})(6{\rm i}-15\cos\theta-6{\rm i}\cos2\theta-\cos3\theta)}{(\cos\theta-{\rm i})(6{\rm i}
 +15\cos\theta-6{\rm i}\cos2\theta+\cos3\theta)}.
\end{equation}
Because of
\begin{equation}
 \lim\limits_{r\to\infty}\frac{r_i}{r}=1,
\end{equation}
see (\ref{ri}), we find (with $\mathcal E_0=\mathcal E_{\rm NHG}$)
\begin{equation}
 \lim\limits_{r\to\infty}
 \frac{\mathcal E}{\mathcal E_{\rm NHG}}=1
\end{equation}
for all $\theta$ with $F(\theta)\neq G(\theta)$ if $n$ of the $2n$ constants $c_i$, say $c_p$ (with $p=1,3,5,\dots, 2n-1$), 
are chosen finite and the other ones, say $c_q$ 
(with $q=2,4,6,\dots, 2n$), are chosen infinite. This reduces the number of free real constants contained in the $K_i$'s and $c_i$'s from $4n$ to $3n$. 
For $n=1$, $c_1\neq\infty$ and $c_2=\infty$ means that $K_1$ and $K_2$ must be real, see (\ref{condci}). For $n>1$, 
pairs of complex conjugate $K_i$'s are possible as well. It turns out that the so far excluded special values of $\theta$ 
defined by $F(\theta) = G(\theta)$ are the same values for which $\alpha_0\equiv-H^*/H=1$ holds ($\cos^2\theta=2\sqrt{3}-3$), leading obviously to
$\mathcal E = \mathcal E_{\rm NHG}$ for all $r$. Hence our solutions have the extreme Kerr near-horizon asymptotics whenever
precisely $n$ of the $2n$ constants $c_i$ are chosen infinite\footnote{Note that these infinite constants can be identified 
with $c_q$ ($q=2,4,6,\dots, 2n$) without loss 
of generality since the expression (\ref{det}) is invariant under simultaneous permutations of the $2n$ indices of the $\lambda_i$ and $\alpha_i$.}.

Explicit expressions for all metric functions can be calculated using the general
B\"acklund formalism, see \cite{kramer80}. For convenience of the reader, the metric functions in the case $n=1$ are given in the Appendix.
%%%%%%%%%%%%%%%%%%%%%%%%%%%%%%%%%%%%%%%%%%%%%%%%%%%%%%%%%%%%%%%%%%%%%%%
\section{Conclusion}
Eqs (\ref{det}, \ref{lambda}) with $\mathcal E_0=\mathcal E_{\rm NHG}$ and (\ref{al0})--(\ref{C}) represent 
a new hierarchy of solutions to the vacuum Einstein equations, the ($2n$-fold) B\"acklund transforms of
the extreme Kerr near-horizon geometry ($n=1,2,3,\dots$). For given $n$, the solution 
contains $4n$ arbitrary real parameters. 
From a mathematical point of view, this result is an example of applying B\"acklund transformations to 
non-static and non-asymptotically flat seed solutions. As discussed in the Introduction, members of the $3n$-parameter subfamily with extreme Kerr 
near-horizon asymptotics might be of particular interest. Of course, further investigations are needed to explore the physical significance of these solutions.  
%%%%%%%%%%%%%%%%%%%%%%%%%%%%%%%%%%%%%%%%%%%%%%%%%%%%%%%%%%%%%%%%%%%%%%%
\section*{Appendix}
For the case $n=1$, the three-parameter\footnote{The three parameters are given by the two real constants $K_1$, $K_2$ and the imaginary 
constant $c_1$ (remember that $c_2=\infty$).} family of solutions with extreme Kerr near-horizon asymptotics 
leads to the following expressions for the metric functions in (\ref{line}):
\begin{equation}
 f=\Re\left(\mathcal E_ {\rm NHG}\; \frac{\alpha_1 r_1-\alpha_2 r_2+\alpha_0(K_1-K_2)}
 {\alpha_1 r_1-\alpha_2 r_2+K_1-K_2}\right),
\end{equation}
\begin{equation}
 h=h_0\Omega^4  K_1^4\psi_1\psi_2\psi_1^*\psi_2^*(\lambda_1\lambda_2^*+\lambda_1^*\lambda_2-\alpha_1\alpha_2^*-\alpha_1^*\alpha_2)f_0^{-2}
\end{equation}
and 
\begin{equation}
 a=(a_0f_0-2\,\Im\, Q)f^{-1},
\end{equation}
where 
\begin{equation}
 h_0=h_{\rm NHG}=\frac{1}{4}(\cos^4\theta+6\cos^2\theta-3),
\end{equation}
\begin{equation}
 f_0=f_{\rm NHG}=\frac{4\Omega^2r^2 h_0}{\cos^2\theta+1},
\end{equation}
\begin{equation}
 a_0=a_{\rm NHG}=-\frac{2r\sin^2\theta}{f_0(\cos^2\theta+1)},
\end{equation}
\begin{equation}
 \psi_1=\psi(\lambda_1,\theta)-c_1\psi(-\lambda_1,\theta), \quad \psi_2=-{\rm i}\psi(-\lambda_2,\theta),
\end{equation}
and
\begin{equation}
 Q=\frac{\alpha_1r_1(K_2-r\cos\theta)-\alpha_2r_2(K_1-r\cos\theta)+{\rm i}a_0f_0(K_1-K_2)}{\alpha_1 r_1-\alpha_2 r_2+K_1-K_2}.
\end{equation}
Note that
\begin{equation}
 \lim\limits_{r\to\infty}\frac{f}{f_0}=\lim\limits_{r\to\infty}\frac{h}{h_0}=\lim\limits_{r\to\infty}\frac{a}{a_0}=1.
\end{equation}
%%%%%%%%%%%%%%%%%%%%%%%%%%%%%%%%%%%%%%%%%%%%%%%%%%%%%%%%%%%%%%%%%%%%%%%%%%%%%%%%%%%%%%%%%%%%%%%%%%%%%%%%%%%%%%

\end{document}